\documentclass[prl,aps,showpacs,twocolumn]{revtex4}
\usepackage{epsfig,color}
\usepackage{bm} 

\renewcommand{\it}[1]{\textit{#1}}
\newcommand{\Onecol} {\begin{widetext} \onecolumngrid} %% 2 -> 1
\newcommand{\Twocol} {\end{widetext} \twocolumngrid} %% 1 -> 2
\newcommand{\be}{\begin{equation}}
\newcommand{\ba}{\begin{array}}
\newcommand{\bea}{\begin{eqnarray}}
\newcommand{\bfi}{\begin{figure}}
\newcommand{\ee}{\end{equation}}
\newcommand{\ea}{\end{array}}
\newcommand{\eea}{\end{eqnarray}}
\newcommand{\efi}{\end{figure}}

\begin{document} 
\title{Stochastic resonance  in a simple model of magnetic reversals}
\author{Roberto Benzi${}^{(1)}$, Jean-Fran\c{c}ois Pinton${}^{(2)}$}
\affiliation{(1) Dip. di Fisica and INFN, Universit\`a ``Tor Vergata"\\
Via della Ricerca Scientifica 1, I-00133 Roma (Italy)\\ 
(2) Laboratoire de Physique, CNRS \& \'Ecole Normale Sup\'erieure de Lyon, \\ 
46 all\'ee d'Italie, F69007 Lyon (France)}

\begin{abstract} 
We discuss the effect of stochastic resonance in a simple model of magnetic reversals. The model exhibits statistically stationary solutions and bimodal distribution of the large scale magnetic field. We observe a non trivial  amplification of stochastic resonance induced by turbulent fluctuations, i.e. the amplitude of the external periodic perturbation needed for stochastic resonance to occur is much smaller than the one estimated by the equilibrium probability distribution of the unperturbed system. We argue that similar amplifications can be observed in many physical systems where turbulent fluctuations are needed to maintain large scale equilibria. 
\vskip 0.2cm 
\end{abstract} 


\maketitle 
%%%%%%%%%%%%%%%%%%%%%%%%%%%%%%%%%%%%%%%%%%%%%%%%%%%%%%%%%%%%%%

\section{The problem}

In this paper we discuss the effect of Stochastic Resonance (SR) for a simple model of magnetic reversals. As we will discuss later on, we show that the effect of SR can be
amplified for systems where statistically stationary states are observed, i.e. where metastable equilibria are due to non linear equilibration between turbulent fluctuations and
non linear large scale effect. We argue that this effect is relevant in many physical systems and could be eventually observed experimentally. 

\bigskip
Before defining more precisely the problem we want to focus, it is worthwhile to review shortly the basic idea behind SR.
SR was  introduced almost 30 years ago  in \cite{benzi} and \cite{nicolis} within the framework of long term climate theory.
One can be understood the mechanism of SR  in the simple case of the stochastic differential equation:
\begin{equation}
\label{1}
d \phi = (m \phi- g\phi^3) dt + \sqrt{\sigma} dW(t)
\end{equation}
where $dW(t)$ is gaussian white noise $\delta-$correlated in time. Because of the noise, $\phi$ shows a bimodal probability distribution peaked around 
 $\pm \phi_m$ where $\phi_m^2 = m/g$. The average transition time $ \tau $ between the two peaks  is proportional to 
\begin{equation}
\tau \sim \exp\frac{2 g \phi_m^2}{\sigma}
\end{equation}
It is well known that the transition time  is a random variable exponentially distributed for small $\sigma$. If we add on the r.h.s of (\ref{1}) a periodic perturbation
$A sin(\omega t)$, something interesting can happen. For $ \pi / \omega  \sim  \tau $ the behavior of $\phi$ becomes nearly periodic, i.e. $\phi$ "jumps" between the two states $\pm \phi_m$ periodically with period $2 \pi/\omega$.  This behavior can be understood in a number of different ways and we refer the reader to the original papers \cite{benzi} and \cite{nicolis}, see also \cite{review} for a review on SR.  For $\omega$ small and the deterministic
time scale $1/m$ much shorter than 
$  \tau $ , the condition
for SR to occur \cite{benzi} can be written as
\begin{equation}
\label{SR}
\frac{A \phi_m}{ \sigma}  \sim 2
\end{equation}
In the simplified model (\ref{1}), the equilibrium probability distribution $P(\phi)$ is peaked around $\pm \phi_m$ which are stable stationary solutions of (\ref{1}) for $\sigma=0$.
In many physical systems, however, the probability distribution of the relevant order parameter (let us still call it $\phi$ ) is bimodal although there exists no stable stationary
states: the peaks in the probability distribution arise because non trivial equilibration of internal dynamics which {\it on the average} can be approximated by an effective equation
similar to (\ref{1}). This implies that the peaks in the bimodal probability distribution should correspond to {\it statistically stationary solutions}. A particular interesting case is the one where statistically stationary solutions are due to the balance between non linear terms and internal fluctuations. In these cases, the effect of an external periodic perturbation can
change the magnitude of the fluctuations which, in turn, change the parameters of the effective equation (\ref{1}) i.e. the value of $\phi_m$. Because of (\ref{SR}), we can observe
an amplification of SR.  Such a mechanism is indeed observed in \cite{benzisutera} for the case of two dimensiona Landau-Ginzburg equation. It is the purpose of this paper to describe the amplification of traditional SR in the case of a simplified model of magnetic reversals. Our major point is that, due to turbulent fluctuations in the presence of 
statistically stationary solutions, SR is strongly amplified. Our example is just one of the many possible cases where the same amplification of SR may be observed and the present study outline  the role of turbulent fluctuations in amplification of SR. We argue that other cases, relevant to turbulent flows and large scale dynamics of geophysical flows,
may show similar amplification.

\bigskip

\section{Simple model of magnetic reversals}

{The question of transitions between statistically solutions is central in the behavior of many out-of-equilibrium systems in physics and geophysics~\cite{atmos,Busse1997,convec,Monchaux07}. As one particular example addressed here, we note that  natural dynamos are intrinsically dynamical.}  Complex magnetic field evolutions have been reported for many systems, including the Sun and the Earth~\cite{RobertsBook}. Formally, the coupled set of momentum and induction equations are invariant under the transform: $(\mathbf{u}, \mathbf{B}) \rightarrow (\mathbf{u}, -\mathbf{B})$ so that states with opposite polarities can be generated from the same velocity field ($\mathbf{u}$ and $\mathbf{B}$ are respectively the velocity and magnetic fields). Such reversals have been observed recently in laboratory experiments using liquid metals, in arrangements where the dynamo cycle is either favored artificially~\cite{BVK} or stems entirely from the fluid motions~\cite{VKSP1,VKSP2}. In these laboratory experiments, as also presumably in the Earth core, the ratio of the magnetic diffusivity to the viscosity of the fluid (magnetic Prandtl number $P_M$) is quite small. As a result, the kinetic Reynolds number $R_V$ of the flow is very high because its magnetic Reynolds number $R_M=R_V P_M$ needs to be large enough so that the stretching of magnetic fields lines balances the Joule dissipation. Hence, the dynamo process develops over a turbulent background and in this context, it is often considered as a problem of `bifurcation in the presence of noise'. 

Building upon the above observations, we shortly review here a recent model proposed in \cite{benzipinton} which 
which incorporates hydromagnetic turbulent fluctuations (as opposed to `noise') in a dynamo instability. 
We  consider an  `energy cascade' model {\it i.e.} a shell model aimed at reproducing few of the relevant characteristic features of the  statistical properties of the Navier-Stokes equations \cite{luca}. In a shell models, the basic variables describing the `velocity field' at scale  $r_n = 2^{-n} r_0 \equiv k_n^{-1}$, is a complex number $u_n$ satisfying a suitable set of non linear equations {(here $r_0=2$)}. There are many version of shell models which have been introduced in literature. Here we choose the one referred to as {\em Sabra} shell model. MHD shell model -- introduced in~\cite{Frick1998} -- allow a description of turbulence at low magnetic Prandtl number since the steps of both cascades can be freely adjusted~\cite{Stepanov2006, Stepanov2007}.
We consider  a formulation extended from the Sabra hydrodynamic shell model: 
\begin{eqnarray}
\frac{du_n}{dt}  & = & \frac{i}{3} (\Phi_n(u,u) - \Phi_n(B,B)) - \nu k_n^2 u_n + f_n\ , \label{1bp}\\
\frac{dB_n}{dt} & = & \frac{i}{3} (\Phi_n(u,B) - \Phi_n(B,u)) - \nu_m k_n^2 B_n   \ , \label{2bp}
\end{eqnarray}
where {$n=1,2,...$ and}
\begin{eqnarray}
&& \Phi_n(u,w) = k_{n+1}[(1+\delta) u_{n+2}w_{n+1}^*+(2-\delta)u_{n+1}^*w_{n+2}]\nonumber\\
&&+ k_{n}[(1-2\delta)u_{n-1}^*w_{n+1}-(1+\delta)u_{n+1}w_{n-1}^*]\nonumber\\
&&+ k_{n-1}[(2-\delta)u_{n-1}w_{n-2}+(1-2\delta)u_{n-2}w_{n-1}] \ ,
\end{eqnarray}
for which  following \cite{benzi05} we chose $\delta= -0.4$. For this value of $\delta$, the {\it Sabra} model is known to show statistical properties (i.e. anomalous scaling) close to the ones observed in  homogenous and isotropic turbulence.  The model, without forcing and dissipation, conserve the kinetic energy {$E_V=$} $\Sigma_n |u_n|^2$, the magnetic energy {$E_B=$} $\Sigma_n |B_n|^2$ and the cross-helicity  $Re(\Sigma_n u_n B_n^*)$. \ In the same limit, the model has a $U(1)$ symmetry corresponding to a phase change $\exp(i\theta)$ in both complex variables $u_n$ and $B_n$. The quantity $\Phi_n(v,w)$ is the shell model version of the transport term $\vec{v}\nabla\vec{w}$. The forcing term $f_n$ is given by $f_n \equiv \delta_{1n}f_0/u_1^*$, {\it i.e.} we force with a constant power injection in the large scale.    We want to introduce in eq. (\ref{2bp}) an extra (large scale) term aimed at producing two statistically stationary equilibrium solutions for the magnetic field. For this purpose, we add to the r.h.s. of (\ref{2bp}) an extra term $M_2(B_2)$,
namely for $n=2$ eq.(\ref{2bp}) becomes:
\begin{equation}
\label{b2}
\frac{dB_2}{dt} = F_2(u,B) - M_2(B_2)-\nu_m k_2^2 B_2
\end{equation}
where $F_2(u,B)$ is a short hand notation for $i/3 (\Phi_2(u,B)-\Phi_2(B,u))$. The term $M_2(B_2)$ is chosen with two requirements: 1) it must break the   $U(1)$ symmetry; 2) it must introduce a large scale dissipation needed to equilibrate the large scale magnetic field production. There are many possible ways to satisfy these two requirements. Here we simply choose $M_2(B_2)=a_m B_2^3$.  From a physical point of view, symmetry breaking also occurs in real dynamos since the magnetic field is directed in one preferential direction which changes sign during a reversal. Also, large scale dissipation must be responsible of the equilibration mechanism of the large scale field. The choice of a non linear equilibration is made here to highlight the the existence of a non linear center manifold for the large scale dynamics. In other words, eq.(4) with $M_2(B_2) = a_m B_2^3$ is supposed to describe the `normal form' dynamics of the large scale magnetic field. Note, that our assumption on $M_2$ does not necessarily imply  a time scale separation between the characteristic time scale of $B_2$ and the magnetic turbulent field. Finally,  since the  system has  an inverse cascade of helicity, we set $B_1=0$ as boundary condition at large scale in order to prevent non stationary
behavior.

The free parameters of the model are the power input  {$f_0$}, the magnetic viscosity $\nu_m$ and the saturation parameters $a_m$.  {Our numerical simulations have been done with  $n=1,2..,25$}. Actually, the parameter {$f_0$} could be eliminated by a suitable rescaling of the velocity field. We shall keep it fixed  to {$f_0 = 1-i$}.  
In this system, a possible estimate of Reynolds numbers is $R_{V} =\sqrt{\langle E_V \rangle}/k_2\nu = \sqrt{\langle E_V \rangle} r_0/4\nu$ and $R_M = \sqrt{\langle E_V \rangle} r_0/4\nu_m$.

For very large $\nu_m$, the magnetic field does not grow. Then, for $\nu_m $ lower than some critical value, $\langle B_2 \rangle$ as well as $E_B$ increases for decreasing $\nu_m$. Eventually, $\langle |B_2| \rangle $ saturates at a given value while $E_B$ still increases, showing that for $\nu_m$ small enough a fully developed spectrum of $B_n$ is achieved. This {type of} behavior  is in agreement with previous {studies of}  Taylor-Green flows~\cite{Ponty05,Laval06}, $s_2t_2$ flows in a sphere~\cite{Bayliss07} or MHD shell models~\cite{Frick06}. 

The onset of dynamo implies that there exists a net flux of energy from the velocity field to the magnetic field.  At the largest scale, the magnetic field $B_2$ is forced by the velocity field due to the terms $F_2(u,B)$. The quantity $S \equiv {\cal R}[F_2(u,B)B_2^*]$ in (\ref{b2}) is the energy pumping due to the velocity field which is independent on $B_2$ and $a_m$. Thus, from eq.(\ref{2}) we can obtain:
\begin{equation}
\label{A}
\frac{1}{2} \frac{d|B_2|^2}{dt} = S - a_m |B_2|^2(B_{2r}^2-B_{2i}^2)- \nu_m k_2^2 |B_2|^2
\end{equation}
where $B_{2r}$ and $B_{2i}$ are the real and imaginary part of $B_2$. For large $\nu_m$, the amplitude of $B_2$ is small and the symmetry breaking term proportional to $a_m$ is negligible. Under this condition, and with the boundary condition constrains, we expect from (\ref{A}) or (\ref{b2}) that the behavior of $B_2$ is periodic, as it has been observed in the numerical simulations. On the other hand for relatively small $\nu_m$, the non linear equilibration breaks the U(1) symmetry and $B_{2i}$ becomes rather small and statistically stationary solutions can be observed with $B_{2r}^2 = \sqrt{S/a_m}$. 
%%%%%%%%%%FIGURA 2%%%%%%%%%%%%%%%%%
\begin{figure}[t]
\centerline{\includegraphics[width=8cm]{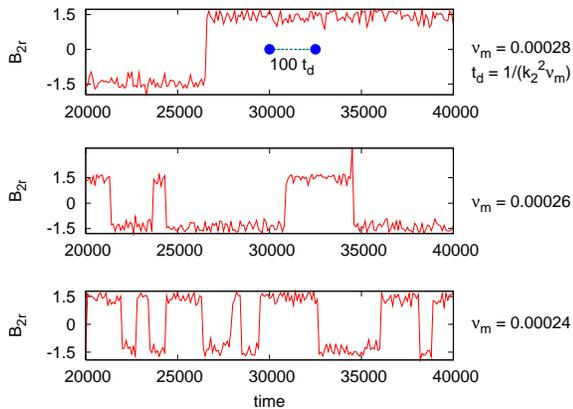}}
\caption{ 
Time behavior of $B_{2r}$ for three different values of $\nu_m$ (displayed on the left side) and constant $\nu$. The blue segment in the upper panel shows $100t_d$, where $t_d$ is the dissipative time scale computed
as $t_d=1./(k_2^2\nu_m)$. One time unit in the figure corresponds to the large scale eddy turnover time $1./(k_1|u_1|)$. {Numerical simulations have run for much longer than the time intervals shown here -- in the complete time series there is no asymmetry in between the $\pm B_2$ states.}
}
\label{fig2}
\end{figure}
%%%%%%%%%%%%%%%%%%%%%%%%%%%%%%%%
In \cite{benzipinton},  it is discussed a systematic study of
the magnetic reversal as a function of $\nu_m$. In figure~\ref{fig2}, we show three different time series of the $B_{2r} = Re(B_2)$ as a function of time for three different, relatively large, values of the magnetic diffusivity. The figure highlights the two major informations, namely the obervation of reversals between the two possible large scale equilibria and the dramatic increase of the time delay between reversals for increasing $\nu_m$  values. Note that this long time scale, as observed in the upper panel of figure~\ref{fig2}, is much longer than the characteristic time scale of $B_2$ near one of the two equilibrium states. The system spontaneously develops a significant time scale separation, for which given polarity is maintained for times much longer  than the magnetic diffusion time. In figure~\ref{fig3} we show the average reversal time as a function of $\nu_m$.  

%%%%%%%%%%%%%%%%%%%%%%%
\begin{figure}[h]
\begin{center}
\includegraphics[width=0.45\textwidth]{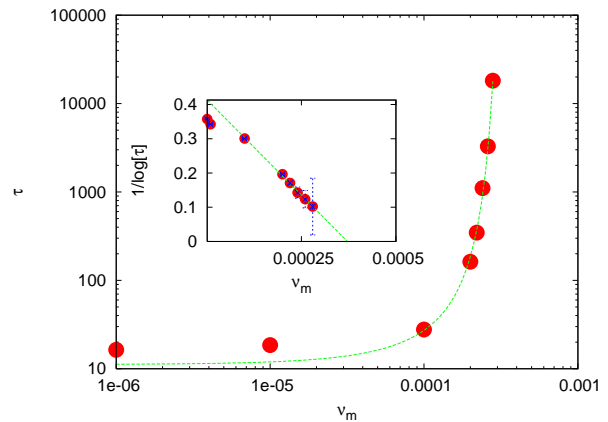}
\end{center}
\caption{Average persistence time $\tau$ as a function of the magnetic viscosity $\nu_m$ for $a_m=0.1$ and $\nu=10^{-7}$. The green line corresponds to the fit given by equation (\ref{fit}). In the insert we plot
{$1/\ln(\tau)$ and its error bars} versus $\nu_m$ to highlight the linear behavior predicted by (\ref{fit}). {Note that the error bars are smaller than the symbol size except for the very last point}
}
	\label{fig3}
\end{figure}
%%%%%%%%%%%%%%%%%%%%%%%%

In order to develop a theoretical framework aimed at understanding the result shown in figure~\ref{fig3},  we assume, in the region where $\langle |B_2|^2 \rangle$ is independent on $\nu_m$, that $B_{2i} \sim 0$ and that the term $F_2(u,B)$ can be divided  into an average forcing term proportional to $B_{2r}$ and a fluctuating part:
\begin{equation}
\label{approx}
F_2(u,B) = \beta B_2 + \phi'
\end{equation}
where $\beta$ depends on {$f_0$} and $\phi'$ is supposed to be uncorrelated with the dynamics of $B_2$ , i.e. $ \langle [\phi' B_2^*] \rangle = 0$. Note that in the context of the mean-field approach to MHD, the first term $\beta B_2$ would correspond to an `alpha-effect'. Using (\ref{approx}), we can rewrite the equations for $B_2$ as follows:
\begin{equation}
\label{b2}
\frac{dB_2}{dt} = \beta B_2 - a_m B_2^3  + \phi' \ .
\end{equation}
where we neglect the dissipative term since $\beta \gg \nu_m k_2^2$ in the region of interest. eq.(\ref{b2}) must be considered an {\it effective} equation describing the dynamics of the magnetic field $B_2$ and its reversals, and the fluctuations $\phi'$ incorporates the turbulent fluctuations from the velocity and magnetic field turbulent cascades. It is the effect of $\phi'$ which makes the system `jump' between the two statistically stationary states. Using (\ref{A}) we can obtain $\beta = \sqrt{S a_m}$ while the two statistical stationary states can be estimated as $\pm B_0$, $B_0^2 = \beta/a_m$. 
% and, using large deviation theory, we can predict $\tau$ to be

%%%%%%%%%%FIGURA 2%%%%%%%%%%%%%%%%%
\begin{figure}[t]
\centerline{\includegraphics[width=8cm]{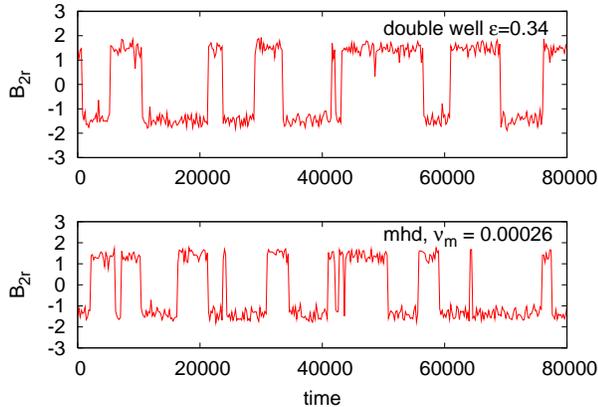}}
\caption{ Solutions of eq. (\ref{2} as compared to the behavior of $B_{2r}$ obtained from (\ref{1bp})-(\ref{2bp}) for $\nu_m=0.00026$. The choice of $\sigma$ in (\ref{2}) is chosen to reproduce the mean transition time $\tau$ close to the solution of (\ref{1bp})-(\ref{2bp}) }

\label{fig4}
\end{figure}
%%%%%%%%%%%%%%%%%%%%%%%%%%%%%%%%

Interpreting eq.(\ref{b2})  as an effective stochastic differential equation, 
 we can predict $\tau$ to be
\begin{equation}
\label{tau}
\tau  \sim \exp\left( \frac{\beta^2}{a_m \sigma} \right) = \exp\left( \frac{S}{\sigma} \right) \ ,
\end{equation}
where $\sigma$ is the variance of the noise $\phi'$ acting on the system. In \cite{benzipinton} it is suggested that  $\sigma = A (\nu_m^* - \nu_m)/uL$ which leads to
\begin{equation}
\label{fit}
\tau \sim \exp\left( \frac{C}{\nu_m^*-\nu_m} \right) \ ,
\end{equation}
where $C$ is a constant independent of $\nu_m$. This functional form is displayed in figure~\ref{fig3}; it agrees remarkably with the observed numerical values of $\tau$ for a rather large range. In the insert of figure~\ref{fig2} we show $1/\log(\tau)$ as a function of $\nu_m$ to highlight the linear behavior predicted by eq.(\ref{fit}). The physical statement represented by (\ref{fit}) is that the average reversal time should show a critical slowing down for relatively large $\nu_m$.  In other words, we expect that fluctuations around the statistical equilibria  increase as $R_M$ increases. The increase of fluctuations may not be monotonic for very large $R_M$, which explains why we are not able to fit the entire range of $\nu_m$ shown in figure~(\ref{fig2}).

\section{The effect of small periodic forcing}
As shown in the previous section, eq.(\ref{b2}) can be considered an effective stochastic differential equation. More precisely, we can approximate the reversals of magnetic field 
$B_{2r}$ by using eq. (\ref{1}) with $ \phi \equiv B_{2r}$ and suitable values of $m$ and $g$ such that $\phi_m$ corresponds to the magnitude of the observed statistically stationary states in the probability distribution of $B_{2r}$. Finally we must choose $\sigma$ in (\ref{1}) such that the average transition time is equal to the one observed for the magnetic reversals (see figure (\ref{fig4})). For our purpose, we choose $\nu_m= 0.00026$ so that the average transition time is order $4000$. 
Then the effective equation (\ref{1}) is
\begin{equation}
\label{2}
d \phi =  \phi (\phi_m^2 - \phi^2) dt + \sqrt{\sigma} dW(t)
\end{equation}
where $\phi_m = 1.51$ and $\sigma = 0.34$ are the numerical values chosen in (\ref{2}) for the dynamics of $\phi$ to be close to the observed numerical behavior of $B_{2r}$.
In figure (\ref{fig4}) we show the behavior of the numerical solution of (\ref{2}) as compared to the time behavior of $B_{2r}$ computed for $\nu_m=0.00026$. Hereafter we refer to eq. (\ref{2}) as "double well" model while we use the term "mhd"  for eq.s (\ref{1bp})-(\ref{2bp}).

\bigskip

%%%%%%%%%%FIGURA 2%%%%%%%%%%%%%%%%%
\begin{figure}[t]
\centerline{\includegraphics[width=8cm]{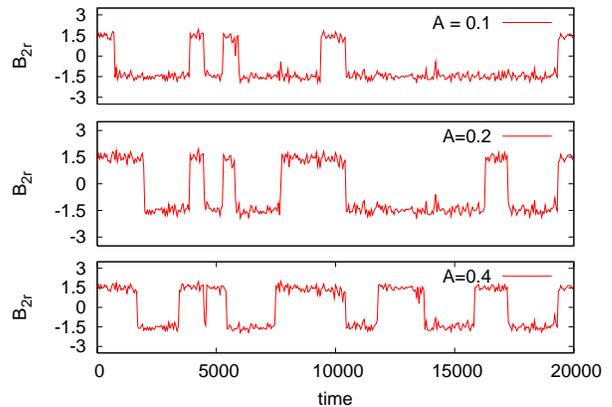}}
\caption{ Numerical solutions of (\ref{3}) for different values of the external periodic forcing $ A sin(2 \pi t/\omega)$. For $A=0.4$ the dynamics of $\phi$ starts to behave periodically}

\label{fig5}
\end{figure}
%%%%%%%%%%%%%%%%%%%%%%%%%%%%%%%%

Following our discussion in the introduction, we can use (\ref{SR}) to estimate the amplitude of an external periodic forcing $Asin(\omega t)$ applied on the r.h.s of (\ref{2})
with $2\pi/\omega = 4000 $.
It turns out that (\ref{SR}) gives $A \sim 0.5$ for SR to occur. This agrees very well with the numerical results shown in figure (\ref{fig5}) where we plot the numerical solution of 

\begin{equation}
\label{3}
d \phi =  \phi (\phi_m^2 - \phi^2) dt + \sqrt{\sigma} dW(t) + A sin(2\pi/\omega t) 
\end{equation}
with $A=0.1$ (upper panel) $A=0.2$ (middle panel) and $A=0.4$ (bottom panel). A more quantitative description can be obtained by looking at the average Fourier amplitude $P_A \equiv \langle |\hat{\phi}(\omega)|\rangle$
and plotting $P_A$ as a function of $A$. This is done in figure (\ref{fig7}) (red circles).
%%%%%%%%%%FIGURA 2%%%%%%%%%%%%%%%%%
\begin{figure}[t]
\centerline{\includegraphics[width=8cm]{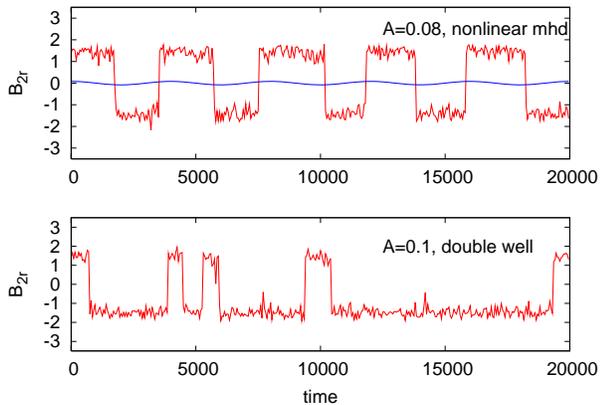}}
\caption{ Upper panel: numerical behavior of $B_{2r}$ obtained from (\ref{1bp})-(\ref{2bp}) when an external forcing $A sin(2\pi t /\omega)$ is added to r.h.s. of (\ref{2bp}) for $n=2$. The blue curve shows the periodic forcing.
Lower panel: solution of (\ref{3}) with $A=0.1$. }

\label{fig6}
\end{figure}
%%%%%%%%%%%%%%%%%%%%%%%%%%%%%%%%
\begin{figure}[t]
\centerline{\includegraphics[width=8cm]{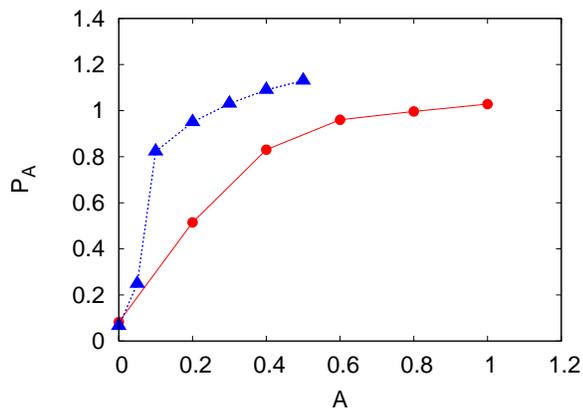}}
\caption{ The average Fourier amplitude $P_A \equiv \langle |\hat{\phi}(\omega)\rangle$ as a function of $A$ for eq. (\ref{3}) (red circles). The blue triangles refer to the same quantity computed for the eq.s (\ref{1bp})-(\ref{2bp}): the response function is much larger than for the case of stochastic differential equation (\ref{3}).
 }

\label{fig7}
\end{figure}
%%%%%%%%%%%%%%%%%%%%%%%%%%%%%%

We now turn our attention to the system (\ref{1bp})-(\ref{2bp}) which describes the magnetic reversals within the simplified model discussed in the previous section.
We add to the equation of motion of $B_{2r}$ an external periodic forcing $A sin(2\pi t /\omega)$ with the same period $2\pi t/\omega = 4000$ chosen for the numerical simulations
of figure (\ref{fig4}). In figure (\ref{fig6}) we show the behavior of $B_{2r}$ as a function of time for $A=0.08$ and compared it against the solution of (\ref{3}) for the same $A$.
We can clearly observe SR in the mhd equations (\ref{1bp})-{\ref{2bp}) while for eq.(\ref{3}) the effect of $A$ is too small. In figure (\ref{fig7}) we show the the average Fourier amplitude $P_A \equiv \langle |\hat{B_{2r}}(\omega)|\rangle$ 
and plotting $P_A$ as a function of $A$ comparing the result with the same quantity computed for the stochastic differential equation (\ref{3}).
The conclusion is that SR is amplified in the mhd equations (\ref{1bp})-(\ref{2bp}) by almost a factor $5$! 

\bigskip
We remark that the effect shown in figure (\ref{fig6}) is higly non trivial, i.e. it is not trivial to figure out an effective stochastic differential equation similar to (\ref{2}) which shows the
same sensitivity to the external perturbation $A$.  Using a different language, borrowed by statistical field theory, we can say the the effective eq. (\ref{2}) is described in terms
of renormalized parameters which depends on the turbulent fluctuations. Since the statistically stationary states are due (on the average) to non linear equilibration between the
fluctuating forcing term $F_2(u,B)$ and the large scale term  $a_m B_2^3$, the effect of external perturbation changes (probably in a non linear way) the amount of fluctuations
which fix the values of the renormalized parameter in (\ref{2}). Another important point to remark is that the behavior observed in our simplified mhd model can be {\it {blindly}} parameterized as "stochastic nose", i.e. although
turbulence is qualitatively acting as a noise in the dynamics, the turbulent fluctuations are correlated to the statistical equilibria in a way which is hardly to parameterize as an external noise.

\section{Conclusion}

In this paper we have shown a relatively simple example of amplification of SR in a system characterized by statistically stationary states. It is interesting to remark that, in our
system, the large scale magnetic field is fluctuating in time around some state which is fully maintained by the turbulent energy flux from the velocity field (i.e. the dynamo instability).
The effect on external perturbation changes this equilibria and amplifies the susceptibility of the system. The amplification observed shows that turbulent fluctuations cannot be parameterized as "noise" {\it{independent}} of statistical equlibria. 

We argue that there may exist many different physical systems where similar effects can be observed. In particular, it will be interesting to explore whether such a strong sensitivity to external perturbation is relevant in geophysical flows where many
theories of large scale multiple equilibria have been proposed in the past. Also, concerning the climate theory, our simple but non trivial example, shows how non linear fluctuations could be coupled to external forcing in a rather non intuitive way. This may open the possibility to reconsider SR in climate theory in the framework of more
complex models where internal turbulent fluctuations of climate dynamics are explicitly taken into account and simulated. Finally, within the application of our present
result in the case of dynamo instability, we argue whether a similar effect could be observed experimentally, i.e. whether by applying a small external forcing a large SR is observed for relatively low value of the external amplitude.

{\bf Acknowledgments}  
This paper has been written for celebrating Madame Catherine Nicolis birthday and her long and outstanding career.  One of the author (RB) have had many occasions in the past to work together with Catherine  and to enjoy many long conversations on how to push
the idea of non trivial dynamic behavior within the framework of stochastic differential equations applied to climate theory. Although the idea was
accepted with some interest by the scientific community working on climate, the basic physical point of representing short term dynamic behavior of climate variables
by using external noises was considered, for many years,  to be more a mathematical curiosity rather than a deep physical intuition. It is a pleasure to recognize that, after 30 years, the scientific community does consider today the effect of noise as physically meaningful and potentially important in both climate theory and weather forecasting. We are honored to share this achievement with Catherine and her husband Gregoire and thank Catherine for her wonderful work.

%\bibliography{/home/biferale/PAPERS/PhysRep/Style/literatur}

\end{document}